\documentstyle[twoside,fleqn,espcrc2,epsfig]{article}

\def\proof{\noindent{\sl Proof:}\kern0.6em}

\def\frac#1#2{\hbox{$#1\over#2$}}
\def\dual{\mathstrut^*\kern-0.1em}

\def\lvec#1{\setbox0=\hbox{$#1$}
    \setbox1=\hbox{$\scriptstyle\leftarrow$}
    #1\kern-\wd0\smash{
    \raise\ht0\hbox{$\raise1pt\hbox{$\scriptstyle\leftarrow$}$}}
    \kern-\wd1\kern\wd0}
\def\rvec#1{\setbox0=\hbox{$#1$}
    \setbox1=\hbox{$\scriptstyle\rightarrow$}
    #1\kern-\wd0\smash{
    \raise\ht0\hbox{$\raise1pt\hbox{$\scriptstyle\rightarrow$}$}}
    \kern-\wd1\kern\wd0}

\def\nabstar#1{\nabla\kern-0.5pt\smash{\raise 4.5pt\hbox{$\ast$}}
               \kern-4.5pt_{#1}}

\def\drvstar#1{\partial\kern-0.5pt\smash{\raise 4.5pt\hbox{$\ast$}}
               \kern-5.0pt_{#1}}

\def\mev{{\rm MeV}}

\def\fm{{\rm fm}}

\def\psibar{\overline{\psi}}

\def\rhoprime{\rho\kern1pt'}
\def\rhobar{\bar{\rho}}
\def\rhobarprime{\rhobar\kern1pt'}
\def\rhobartilde{\kern2pt\tilde{\kern-2pt\rhobar}}
\def\rhobartildeprime{\kern2pt\tilde{\kern-2pt\rhobar}\kern1pt'}

\def\zetabar{\bar{\zeta}}
\def\zetaprime{\zeta\kern1pt'}
\def\zetabarprime{\zetabar\kern1pt'}
\def\zetar{\zeta_{\raise-1pt\hbox{\sixrm R}}}
\def\zetabarr{\zetabar_{\raise-1pt\hbox{\sixrm R}}}

\def\phiimpr{\phi_{\kern0.5pt\hbox{\sixrm I}}}

\def\diracstar#1#2{
    \setbox0=\hbox{$\gamma$}\setbox1=\hbox{$\gamma_{#1}$}
    \gamma_{#1}\kern-\wd1\kern\wd0
    \smash{\raise4.5pt\hbox{$\scriptstyle#2$}}}

\def\ba{b_{\rm A}}
\def\bv{b_{\rm V}}

\def\bm{b_{\rm m}}

\def\ca{c_{\rm A}}
\def\cv{c_{\rm V}}
\def\csw{c_{\rm sw}}

\def\f1{f_1}

\def\opprime#1{\setbox0=\hbox{${\cal O}$}\setbox1=\hbox{${\cal O}_{\rm #1}$}
    {\cal O}_{\rm #1}\kern-\wd1\kern\wd0
    \smash{\raise4.5pt\hbox{\kern1pt$\scriptstyle\prime$}}\kern1pt}

\def\ophatprime#1{\setbox0=\hbox{$\widehat{\cal O}$}
    \setbox1=\hbox{$\widehat{\cal O}_{\rm #1}$}
    \widehat{\cal O}_{\rm #1}\kern-\wd1\kern\wd0
    \smash{\raise4.5pt\hbox{\kern1pt$\scriptstyle\prime$}}\kern1pt}

\def\bopprime#1{\setbox0=\hbox{${\cal O}$}\setbox1=\hbox{${\cal O}_{\rm #1}$}
    {\cal L}_{\rm #1}\kern-\wd1\kern\wd0
    \smash{\raise4.5pt\hbox{\kern1pt$\scriptstyle\prime$}}\kern1pt}

\def\blagprime#1{\setbox0=\hbox{${\cal B}$}\setbox1=\hbox{${\cal B}_{#1}$}
    {\cal B}_{#1}\kern-\wd1\kern\wd0
    \smash{\raise5.2pt\hbox{\kern1pt$\scriptstyle\prime$}}\kern1pt}

\def\mq{m_{\rm q}}
\def\mqtilde{\widetilde{m}_{\rm q}}

\def\za{Z_{\rm A}}

\def\zv{Z_{\rm V}}

\def\msbar{{\rm \overline{MS\kern-0.05em}\kern0.05em}}

\def\kc{\kappa_{\rm c}}
\def\fpsr{(f_{\rm PS})_{\rm R}}
\def\fps0{f_{\rm PS}^{(0)}}
\def\gps0{g_{\rm PS}^{(0)}}
\def\mps{m_{\rm PS}}
\def\mv{m_{\rm V}}

\title{%
\vspace{-3.1cm}
\begin{flushleft}
       {\normalsize OUTP--97--51--P}    \\[-0.2cm]
       {\normalsize October 1997}   \\
\end{flushleft}
       \vspace{0.7cm}
Verification of O($a$) improvement\thanks{Plenary talk
 presented at ``Lattice 97'', Edinburgh, 22--26 July 1997.}}

\author{Hartmut Wittig\address{Theoretical Physics, Oxford University,
			       1~Keble Road, Oxford OX1~3NP, UK}}

\begin{document}
\newcommand{\ewxy}[2]{\setlength{\epsfxsize}{#2}\epsfbox[30 30 640 640]{#1}}
\newcommand{\be}{\begin{equation}}
\newcommand{\ee}{\end{equation}}
\newcommand{\bea}{\begin{eqnarray}}
\newcommand{\eea}{\end{eqnarray}}
\newcommand{\np}{non-perturbative}
\newcommand{\nply}{non-perturbatively}
\newcommand{\stg}{\sqrt{\sigma}}
\newcommand{\gtaeq}{\raisebox{-.6ex}{$\stackrel{\textstyle{>}}{\sim}$}}
\newcommand{\lesssim}{\raisebox{-.6ex}{$\stackrel{\textstyle{<}}{\sim}$}}

\begin{abstract}
  The status of simulations using the non-perturbatively O($a$)
  improved Wilson action in the quenched approximation is
  reviewed. The impact of non-perturbative improvement on the hadronic
  mass spectrum and the size of residual lattice artefacts in spectral
  quantities and decay constants are assessed.
\end{abstract}
\maketitle

\section{INTRODUCTION}

One of the most important systematic effects in lattice simulations of
QCD is the finiteness of the lattice spacing~$a$. It is well known
that physical observables computed using the Wilson action are subject
to corrections of order~$a$, which can be rather large. In order to
obtain reliable results in the continuum limit it is desirable to
reduce lattice artefacts, either through the Symanzik improvement
programme\,\cite{SymanzikI,SymanzikII}, or by employing a
renormalisation group approach\,\cite{HasNied94,Has_lat97}. Recently
the ALPHA Collaboration has carried out the Symanzik improvement
programme to leading order through a non-perturbative determination of
the O($a$) improved fermion action and isospin
currents\,\cite{Alpha_lat96,paperIII,paperIV,marco_lat97}. This
approach should lead to the complete removal of lattice artefacts of
order~$a$ in spectral quantities and matrix elements of local
currents, so that the remaining cutoff effects are of order~$a^2$. The
\nply\ improved action has already been employed in a number of
simulations in the quenched
approximation\,\cite{QCDSF_97,par_lat97,mendes_lat97}.  Here, we
assess the impact of \np\ improvement on the calculation of the mass
spectrum and decay constants in the light hadron sector, analyse the
scaling behaviour and estimate the size of residual lattice artefacts.

The general expression of the O($a$) improved fermion action
reads\,\cite{SW85} 
\bea
S_F^I[U,\psibar,\psi] & = & S_F^W[U,\psibar,\psi] \nonumber\\
 & & \hspace{-1.5cm}+\,\csw{\textstyle\frac{ia}{4}}\sum_{x,\mu,\nu}
       \psibar(x)\sigma_{\mu\nu}F_{\mu\nu}(x)\psi(x),
\label{EQ:currents}
\eea
where $S_F^W$ is the (unimproved) Wilson fermion action and $\csw$ is
an improvement coefficient. The improved and renormalised axial and
vector currents are defined as\,\cite{paperI}
\bea
(A_{\rm R})_\mu^a\hspace{-0.2cm} & = &\hspace{-0.2cm}
 \za(1+\ba a\mq)\big\{A_\mu^a+\ca a\partial_\mu
P^a\big\},  \\
(V_{\rm R})_\mu^a\hspace{-0.2cm} & = &\hspace{-0.2cm}
 \zv(1+\bv a\mq)\big\{V_\mu^a+\cv a\partial_\nu
T_{\mu\nu}^a\big\},
\eea
where $\za, \zv$ are the renormalisation factors of the respective
currents, and $\ba,\ca,\bv$ and $\cv$ are further improvement
coefficients. The unimproved currents and densities are defined as
\bea
A_\mu^a & = & \psibar\gamma_\mu\gamma_5\frac{\tau^a}{2}\psi,\quad
~V_\mu^a = \psibar\gamma_\mu\frac{\tau^a}{2}\psi,\nonumber\\
P^a & = & \psibar\gamma_5\frac{\tau^a}{2}\psi,\qquad
T_{\mu\nu}^a = i\psibar\sigma_{\mu\nu}\frac{\tau^a}{2}\psi,
\eea
where $\tau^a$ are the Pauli matrices acting in flavour space.
The normalisations $\za,\zv$ and improvement coefficients $\csw, \ca,
\cv, \bv$ have been determined \nply\ for bare couplings $g_0$ in the
range $0\leq g_0\leq1$\,\cite{paperIII,paperIV,marco_lat97}. Results
and proposals for non-perturbative determinations of $\ba$ at
$g_0\simeq1$ have been
reported\,\cite{MRSSTT97,giulia_lat97,ssharpe_lat97}. Furthermore, all
of the above improvement coefficients have been calculated in
perturbation theory to one-loop order\,\cite{paperII,paperV}.

The parameters of the simulations discussed here are listed in
Table\,\ref{Tab:simpar}. In addition to the \np\ value of $\csw$, the
QCDSF collaboration have also used unimproved Wilson fermions for a
direct comparison\,\cite{QCDSF_97}. UKQCD have also used the tadpole
improved value of $\csw$ on the same set of configurations and at
$\beta=5.7$\,\cite{UKQCD_lat96,tadpole_comp}, where a \nply\
determined value of $\csw$ is not available. Table\,\ref{Tab:simpar}
shows that all collaborations have used the same two values of
$\beta$. This implies that one cannot test as yet whether \np\
improvement indeed leads to a scaling behaviour of physical
observables which is consistent with O($a^2$) corrections. However, if
one {\it assumes\/} this to be the leading scaling behaviour after
improvement, one can still estimate the size of residual lattice
artefacts at a given value of~$a$.

\begin{table}[tb]
\begin{center}
\caption{Simulations using \np\	improvement. The number of exceptional
configurations discarded from the ensemble is shown in brackets.}
\begin{tabular}{lccr}
\hline
Collab.	& $\beta$  & $L^3\cdot T$  & Statistics \\
\hline
QCDSF\,\cite{QCDSF_97}
	& 6.0	   & $16^3\cdot48$ & $\sim1000$~~~	\\
	& 	   & $24^3\cdot48$ & $\sim200$~~~	\\
	& 6.2	   & $24^3\cdot48$ & $\sim300$~~~	\\
\hline
UKQCD\,\cite{par_lat97}
	& 6.0      & $16^3\cdot48$ & 497(3)	\\
	&	   & $32^3\cdot64$ &  70(1)	\\
	& 6.2	   & $24^3\cdot48$ & 251~~~	\\
\hline
APETOV\,\cite{mendes_lat97}
	& 6.0	   & $16^3\cdot48$ &  50(1)	\\
	& 6.2	   & $24^3\cdot48$ &  50~~~~	\\
\hline
\end{tabular}
\end{center}
\label{Tab:simpar} 
\vspace{-0.8cm}
\end{table}

\section{SPECTRAL QUANTITIES}

\begin{figure}[tb]
\vspace{-1.0cm}
\ewxy{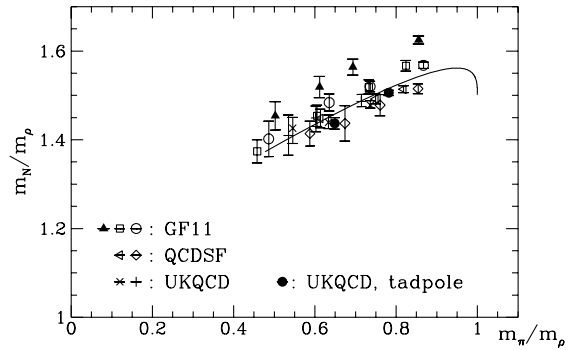}{90mm}
\vspace{-4.5cm}
\caption{Edinburgh plot.}
\label{Fig:edin}
\end{figure}

In Fig.\,\ref{Fig:edin} we show the Edinburgh plot. In addition to the
\nply\ improved data, we also display the tadpole improved results by
UKQCD at $\beta=5.7$ and the data by GF11 using the unimproved action
at $\beta=5.7, 5.93, 6.17$\,\cite{GF11_spec}. One observes that
improvement yields consistently lower values for $m_N/m_\rho$ compared
to the unimproved action. The most dramatic effect is observed at
$\beta=5.7$ when one compares the unimproved results (full triangles)
to the tadpole improved ones (full circles). In fact, it seems that
the mass behaviour obtained using tadpole improvement at $\beta=5.7$
is indistinguishable from the \nply\ improved action. However, as we
shall see later, the residual lattice artefacts in the tadpole
improved data at $\beta=5.7$ are still large. Thus, the Edinburgh plot
disguises rather than exposes lattice artefacts and should therefore
not be used to draw conclusions about the scaling behaviour.

\begin{figure}[tb]
\vspace{-1.0cm}
\ewxy{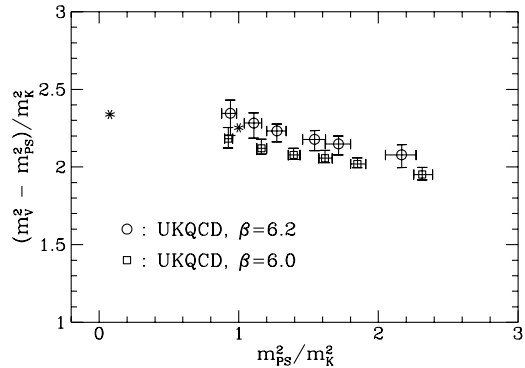}{90mm}
\vspace{-4.3cm}
\caption{UKQCD data for the vector-pseudo-scalar mass splitting
in units of $m_K^2$. Asterisks denote the experimentally observed
values.} 
\vspace{-0.3cm}
\label{Fig:splitting}
\end{figure}

It is a well-known fact that lattice simulations using unimproved
actions fail to reproduce the experimentally observed behaviour of the
vector-pseudoscalar mass splitting, i.e. $\mv^2-\mps^2\sim {\rm
const}$, which holds up to the mass of the charm quark. This is
usually ascribed to the influence of lattice artefacts in the
computation of these splittings. Data for $\mv^2-\mps^2$ by
UKQCD\,\cite{par_lat97} and QCDSF\,\cite{QCDSF_97} using the
\nply\ improved action show that the splittings are close to the
experimental values for the $(\rho,\pi)$ and $(K^*,K)$ systems (see
Fig.\,\ref{Fig:splitting} for a plot of the UKQCD data). Furthermore,
by comparing the results at $\beta=6.0$ and~6.2 one observes that the
dependence on the lattice spacing is small. However, for quark masses
above $m_{\rm strange}$ the two collaborations see a slight downward
trend in the data, so that one can expect that the $D^*-D$ splitting
is still not reproduced correctly.

We now discuss the chiral limit and the critical value of the hopping
paramter, $\kc$. Usually $\kc$ is defined at the point where the
pseudoscalar mass vanishes, $\mps=0$. In accordance with the quark
mass behaviour of $\mps^2$ implied by the PCAC relation, one can
determine $\kc$ from a linear fit to
\be
  (a\mps)^2 = a^2B\frac{1}{2}\left(\frac{1}{\kappa}-\frac{1}{\kc}
  \right) = a^2B\,\mq.
\label{EQ:mps2}
\ee
Both QCDSF and UKQCD have reported that this linear ansatz in
$1/\kappa$ results in poor fits with large correlated $\chi^2/{\rm
dof}$ and have therefore resorted to using model functions which also
contain quadratic terms in $1/\kappa$. However, in order to be
consistent with O($a$) improvement, the quark mass $\mq$ in
eq.\,(\ref{EQ:mps2}) should be replaced by
\be
\mqtilde=\mq(1+\bm a\mq),
\ee
where $\bm=-\frac{1}{2}-0.0962g_0^2$ in one-loop perturbation
theory\,\cite{paperV}. This modification of the fitting ansatz in the
determination of $\kc$ has so far not been used.

Another method defines $\kc$ at the point where the quark mass defined
through the PCAC relation vanishes, i.e. $m_{\rm PCAC}=0$. Here it is
important to realise that 
even after \np\ improvement
chiral symmetry is only approximately restored, so that
\bea
  &\partial_\mu &\hspace{-0.3cm} \big\{A_\mu(x)+\ca a\partial_\mu P(x)\big\}
	\nonumber\\
 & &\hspace{1.0cm} =\hspace{0.1cm} 2m_{\rm PCAC}P(x) + {\rm O}(a^2).
\eea
Therefore, the values of $\kc$ determined by requiring either
$\mps^2=0$ or $m_{\rm PCAC}=0$ will differ by terms of order
$a^2$. The compilation of results in Table\,\ref{Tab:kappac} shows
that in the range of $\beta$ under study the difference in $\kc$ using
either method is of the order of $10^{-4}$ and thus statistically
significant.

\begin{table}[tb]
\begin{center}
\caption{Values for $\kc$ determined at $\mps^2=0$ and at $m_{\rm
PCAC}=0$.} 
\begin{tabular}{cr@{.}lr@{.}lr@{.}l}
\hline
\hline
	& \multicolumn{6}{c}{$\mps^2=0$} \\
\cline{2-7}
$\beta$ & \multicolumn{2}{c}{UKQCD}  & \multicolumn{2}{c}{QCDSF}
        & \multicolumn{2}{c}{APETOV} \\
\hline
6.0     & 0&135335${}^{+20}_{-17}$  & 0&13531(1)
	& \multicolumn{2}{c}{--} \\[0.1cm]
6.2     & 0&135895${}^{+14}_{-55}$  & 0&13589(2)  & 0&135861(19) \\[0.05cm]
\hline
\hline
	& \multicolumn{6}{c}{$m_{\rm PCAC}=0$} \\
\cline{2-7}
$\beta$ & \multicolumn{3}{c}{ALPHA}  & \multicolumn{3}{c}{APETOV} \\[0.05cm]
\hline
6.0	& \multicolumn{3}{c}{0.135196(14)}
	& \multicolumn{3}{c}{--}	\\[0.1cm]
6.2	& \multicolumn{3}{c}{0.135795(13)}
	& \multicolumn{3}{c}{0.135802(6)}	\\[0.05cm]
\hline
\hline
\end{tabular}
\end{center}
\label{Tab:kappac}
\vspace{-0.8cm}
\end{table}

\begin{figure}[tb]
\vspace{-1.0cm}
\ewxy{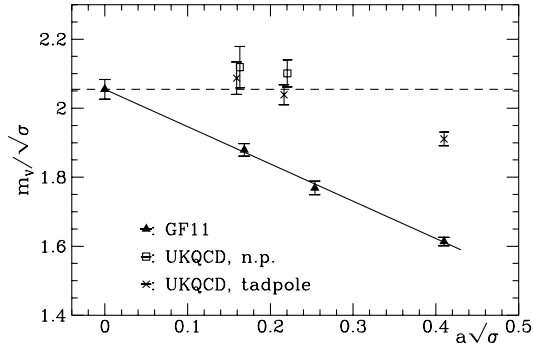}{90mm}
\vspace{-4.5cm}
\caption{Scaling behaviour of the vector mass at $\mps=m_K$. The
dashed line indicates the extrapolated result using the GF11 data.}
\label{Fig:mKstar}
\end{figure}

We now analyse the scaling behaviour of the mass of the vector meson
by comparing the approach to the continuum limit for unimproved and
improved actions. To this end, we note that the continuum limit should
be studied for constant physical volume, so that finite-volume effects
do not distort the scaling behaviour. Since chiral extrapolations are
poorly understood, we will use the available lattice data for the
vector and pseudoscalar masses $a\mv, a\mps$ as well as the string
tension $a\stg$ in order to interpolate the dimensionless ratio
$\mv/\stg$ to $\mps/\stg=1.125$. For $\stg=440\,\mev$ this implies
$\mps=m_K=495\,\mev$. If $\mps^2$ and $\mv$ are linear functions of
the quark masses then one expects that $\mv\simeq m_{K^*}=892\,\mev$
or $\mv/\stg\simeq2.027$. In Fig.\,\ref{Fig:mKstar} we have plotted
the interpolated data for $\mv/\stg$ versus $a\stg$ for
unimproved\,\cite{GF11_spec} and improved
actions\,\cite{par_lat97,UKQCD_lat96,tadpole_comp}. For the unimproved
action one observes that $\mv\simeq m_{K^*}$ is satisfied, but only
after the extrapolation to the continuum limit. In contrast, the data
obtained using the \nply\ or tadpole improved actions at $\beta=6.0,
6.2$ show very little dependence on~$a$ and are rather close to
$\mv\simeq m_{K^*}$. This, however, can no longer be claimed for the
tadpole improved data at $\beta=5.7$, where instead one observes large
residual lattice artefacts. From the slope of the linear fit to
$\mv/\stg$ for the unimproved data, we infer the size of residual
lattice artefacts at $a\simeq0.1\,\fm$ to be $(12\pm1)\%$. For the
\nply\ improved data we estimate the leading corrections of O($a^2$)
to be only $\simeq2\%$.

\begin{figure}[tb]
\vspace{-1.0cm}
\ewxy{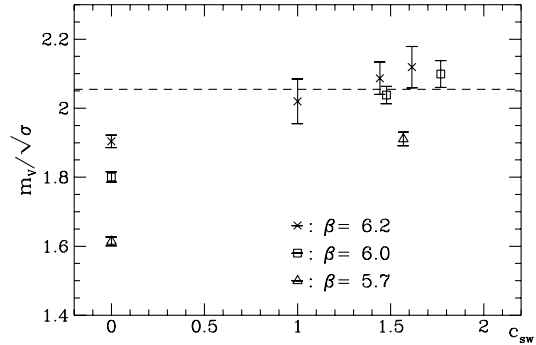}{90mm}
\vspace{-4.5cm}
\caption{$\mv/\stg$ as a function of $\csw$. The sets of
points from right to left have been obtained using \np, tadpole,
tree-level and no improvement. Data at $\beta=5.7$ are shown only for
unimproved and tadpole improved actions. The dashed line shows the
extrapolated GF11 data.}
\label{Fig:mK_vs_csw}
\end{figure}

We can now turn the tables and ask how well various prescriptions to
fix $\csw$ satisfy the ``phenomenological'' improvement condition
\be
{\mv}/{\stg}={\rm const}
\label{EQ:phenimp}
\ee
Besides the data obtained using tadpole and \np\ estimates for $\csw$,
one can use further spectrum data obtained using
$\csw=0$\,\cite{GF11_spec} and $\csw=1$\,\cite{strange}. In
Fig.\,\ref{Fig:mK_vs_csw} we plot $\mv/\stg$ at $\mps=m_K$ versus
$\csw$. From the plot one infers that the condition (\ref{EQ:phenimp})
is satisfied within statistical errors if
\be
   \csw\,\,\gtaeq\,\, c_{\rm sw}^{\rm tadpole}\equiv
   u_0^{-3},\qquad\beta\,\,\gtaeq\,\,6.0,
\label{EQ:cswtad}
\ee
where $u_0$ denotes the average link variable. Thus, with the present
statistical accuracy, already the tadpole improved estimate of $\csw$
leads to a large reduction of O($a$) effects, provided the lattice
spacing is not too large. Fig.\,\ref{Fig:mK_vs_csw} also shows that in
order to satisfy (\ref{EQ:phenimp}) at $\beta=5.7$ one would have to
go to much larger values of $\csw$ than those implied by the tadpole
prescription. This, however, appears impossible due to the appearance
of exceptional configurations\,\cite{paperIII}.

\section{DECAY CONSTANTS}

From the definitions of the renormalised and improved currents,
eq.\,(\ref{EQ:currents}), one sees that in order to obtain O($a$)
improved matrix elements of axial and vector currents one requires
knowledge of the improvement coefficients $\ca,\ba,\cv,\bv$ and the
renormalisation factors $\za$ and $\zv$. In this section we will focus
on the pseudoscalar decay constant $\fpsr$, which we write as
\be
  \fpsr = \za(1+\ba a\mq)\big\{\fps0+\ca \gps0\big\}, 
\ee
where the unrenormalised matrix elements are parametrised as
\bea
  \mps\fps0 & = & \langle0|A_4(0)|{\rm PS}\rangle \\
      \gps0 & = & \langle0|P(0)|{\rm PS}\rangle.
\eea
A major advantage of \np\ improvement is that the renormalisation
factor $\za$ has been determined with an accuracy of around
1\%. At $\beta=6.0$ and 6.2 one finds\,\cite{paperIV}
\be
  \za=\left\{\begin{array}{lr}
	0.791(9), & \quad\beta=6.0 \\
	0.807(8), & \quad\beta=6.2 
	     \end{array}
      \right.
\ee
Note that the error on $\za$ must be combined with the statistical
error on $\fps0+\ca \gps0$.

The improvement coefficient $\ca$ has also been determined
\nply\,\cite{paperIII}:
\be
  \ca=\left\{\begin{array}{lr}
        -0.083, & \quad\beta=6.0 \\
        -0.037, & \quad\beta=6.2
             \end{array}
      \right.
\ee
The effect of $\ca$ on the decay constant can be studied by comparing
$\fps0$ with $\fps0+\ca\gps0$. Here the contribution from $\ca\gps0$
leads to a decrease in the pseudoscalar decay constant of $\sim4\%$ at
$\beta=6.2$ and even $\sim15\%$ at $\beta=6.0$. This effect is
particularly pronounced for $\beta<6.2$.

The improvement coefficient $\ba$ is relevant for the determination of
the kaon decay constant $f_K$. However, unlike $\bv$, the coefficient
$\ba$ has so far not been determined \nply. In one-loop perturbation
theory one finds\,\cite{paperV}
\be
  \ba=1+0.1522g_0^2+O(g_0^4).
\ee
In order to study the influence of $\ba$, one can evaluate $\fpsr$
around the strange quark mass for different choices of $\ba$. Here we
compare
\begin{itemize}
\item $\ba=0$
\item $\ba = 1 + 0.1522g_0^2$
\item $\ba=\bv$
\end{itemize}
The seemingly {\it ad hoc\/} choice of $\ba=\bv$ (here we use the \np\
determination of $\bv$) is motivated by the observation that the
one-loop coefficients for $\ba$ and $\bv$ are approximately
equal\,\cite{paperV}. By applying different choices of $\ba$ to the
analysis of the UKQCD data around $m_K$, one finds that $\ba$ leads to
an increase in $\fpsr$ of at most $\sim2\%$ at $\beta=6.2$ and
$\sim3\%$ at $\beta=6.0$, which is fairly small. Since the choice
$\ba=\bv$ gives essentially the same mass behaviour compared
to choosing $\ba = 1 + 0.1522g_0^2$, one concludes that perturbative
estimates of $\ba$ are quite acceptable for quark masses up to and
around $m_{\rm strange}$.

\begin{figure}[tb]
\vspace{-1.0cm}
\ewxy{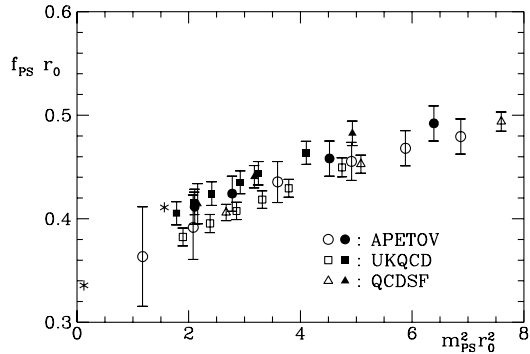}{90mm}
\vspace{-4.5cm}
\caption{The pseudoscalar decay constant as a function of $\mps^2$ in
units of $r_0$ for $\beta=6.0$ (open symbols) and $\beta=6.2$ (full
symbols). Asterisks denote the experimental values for $f_\pi$ and
$f_K$ using $r_0=0.5\,\fm$.} 
\label{Fig:fpsvsmps2}
\end{figure}

\begin{figure}[tb]
\vspace{-1.0cm}
\ewxy{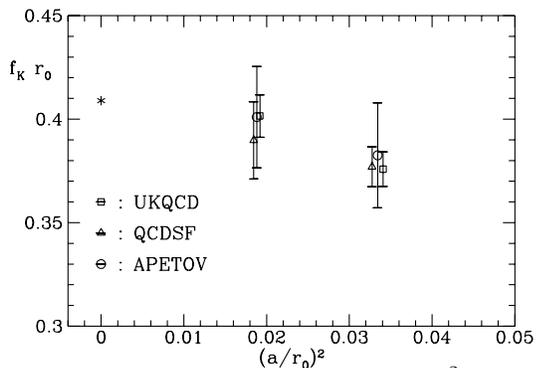}{90mm}
\vspace{-4.5cm}
\caption{$f_Kr_0$ plotted versus $(a/r_0)^2$. The asterisk denotes the
experimental result.}
\label{Fig:fK}
\end{figure}

We now analyse the scaling behaviour of $\fpsr$. In
Fig.\,\ref{Fig:fpsvsmps2} we plot $\fpsr r_0$ versus $(\mps r_0)^2$,
where $r_0$ is the hadronic radius defined in\,\cite{rainer_r0}. Data
for $r_0/a$ were taken from\,\cite{how_r0}. If lattice effects are
small the data in Fig.\,\ref{Fig:fpsvsmps2} should lie on a universal
curve. The results for $f_Kr_0$ computed at $\beta=6.0$ and~6.2 show a
slight dependence on the lattice spacing. In order to study residual
lattice artefacts, we employ a similar procedure as in the case of the
vector mass and extrapolate $\fpsr r_0$ to $(m_K
r_0)^2$. Fig.\,\ref{Fig:fK} shows the resulting values of $f_Kr_0$ as
a function of $(a/r_0)^2$. There is good agreement between the data
from all three collaborations, and in principle their results could be
combined. Furthermore, it appears that a linear extrapolation in $a^2$
yields a continuum result which is compatible with the experimentally
observed value (although there is {\it a priori\/} no reason why the
quenched approximation should reproduce the measured result). In
contrast, the authors of\,\cite{GF11_decay} have found that, for the
unimproved action, the continuum value of $f_K$ is significantly lower
than the experimental result.

From the slope in $(a/r_0)^2$ one estimates that residual lattice
artefacts in $f_K r_0$ amount to ca. 10\% at
$a\simeq0.1\,\fm$. Compared to the previously discussed case of the
vector meson, this is a fairly large correction. Given the substantial
contribution of $\ca\gps0$ to $\fpsr$ at $\beta=6.0$, one can ask
whether $\ca$ has a large effect on the scaling behaviour. If one
formulates a similar improvement condition for $\ca$ as in
eq.\,(\ref{EQ:phenimp}), for instance
\be
   f_K r_0 = {\rm const.}
\ee
one can study how well it is satisfied for different choices of
$\ca$. Using the UKQCD data it turns out that the above condition
``favours'' smaller values of $\ca$ at the lower end of the
$\beta$-range. However, at this stage one should not jump to
conclusions before a more thorough investigation of improvement
conditions has been performed.

\section{CONCLUSIONS}

First results from simulations using the \nply\ improved Wilson action
and currents in the quenched approximation show that lattice artefacts
are drastically reduced. Unlike the case of unimproved actions, the
results for the vector meson mass are practically independent of~$a$
at $\beta=6.0$ and~6.2, so that residual~$a^2$ effects are around
$2\%$ at $a\simeq0.1\,\fm$. 
A real test of the scaling behaviour is still lacking and will only
become possible when more and different values of the lattice spacing
are studied.

The analysis of data for the pseudoscalar decay constant has shown
that a non-perturbative determination of $\ba$ is required for quark
masses above $m_{\rm strange}$. The improvement coefficient $\ca$ has
a large influence on the scaling behaviour, which motivates further
investigation. On the whole, it appears that \np\ improvement leads to
better agreement between the continuum result for $f_K$ and the
experimental value.

The systematic nature of \np\ improvement makes it easily applicable
to other situations: results for $\csw$ computed for two flavours of
dynamical quarks have been reported\,\cite{Karl_Rainer_lat97}.
Furthermore, the formalism has been applied to quenched QCD with an
improved gauge action\,\cite{KlasEd_lat97}.

I thank P. Rowland for help with the UKQCD data and also M. Guagnelli,
T. Mendes and G. Schierholz for communicating their data prior to
publication. I wish to thank M. L\"uscher and R. Sommer for many
helpful discussions.

\end{document}